\documentclass[aps,prd,amsfonts,nofootinbib,supercriptaddress]{revtex4}
\usepackage{amssymb,amsmath}
\usepackage{graphicx}
\usepackage{epsfig}
\usepackage{feynmp}

\newcommand{\beq}{\begin{equation}}
\newcommand{\eeq}{\end{equation}}
\newcommand{\bea}{\begin{eqnarray}}
\newcommand{\eea}{\end{eqnarray}}
\newcommand{\vc}[1]{{\textbf{#1}}}

\begin{document}

\hfill TTK-11-11

\title{Gauge invariance and non-Gaussianity in Inflation}

\author{Gerasimos Rigopoulos}
\affiliation{Institute for Theoretical Particle Physics and Cosmology, RWTH Aachen, D - 52056, Germany}

\begin{abstract} We clarify the role of gauge invariance for the computation of quantum non-Gaussian correlators in inflation. A gauge invariant generating functional for n-point functions is given and the special status of the spatially flat gauge is pointed out. We also comment on the relation between gauge transformations, field redefinitions, the choice of $t={\rm const}$ hypersurfaces and the use of boundary terms in computations of non-Gaussianity.
\end{abstract}
\maketitle

Cosmological Inflation has proven to be a very successful paradigm for the description of the early Universe. Although initially proposed \cite{Guth:1980zm} as an explanation of the gross features of the universe, its spatial flatness, homogeneity etc\footnote{See however \cite{Goldwirth:1991rj} for the sensitivity of inflation on the initial conditions}, its great success has undoubtedly been the explanation of the primordial fluctuations which seeded the cosmic structures observed today. Inflation's predictions for the fluctuations - almost scale invariant and almost Gaussian - fit our best observations remarkably well \cite{Komatsu:2010fb}. The consolidation of the inflationary picture for the early universe and its confrontation with observations has extensively relied on the study of the linearized perturbations generated during inflation and evolved until the emission of the Cosmic Microwave Background and, much later, the onset of cosmic structure formation.

After three decades of progress on the application of linearized perturbation theory, the increasing precision of cosmological data now compels us to study the next to leading order non-linear corrections and their associated observational signature: non-Gaussianity. Indeed, over the past few years non-Gaussianity has emerged as a very fine probe of the inflationary Paradigm, and the early Universe in general, as it can distinguish among different models which are observationally identical on the level of linearized perturbations. Furthermore, the study of the interactions among the different fourier modes of cosmological perturbations during inflation and afterwards, is crucial for the advancement of our theoretical understanding and the demonstration of the self-consistency of the leading order calculations. For these reasons the study of non-linearities and non-Gaussianity has become a very active subfield of current cosmological research (see eg \cite{CQG} for recent reviews and more extensive references).

A large number of works on the calculation of primordial non-Gaussianity from inflation have so far focused on solving some form of the classical equations of motion, perhaps in some gauge invariant formulation, at second order in perturbations with the initial conditions provided after horizon crossing by stochastic classical fields. However, the origin of inflationary fluctuations is ultimately quantum mechanical and a complete calculation would involve the application of techniques from quantum field theory. This is the approach we will pursue here. To be concrete, let us focus on inflation driven by a single scalar field $\Phi$ slowly rolling down the slope of its potential $V(\Phi)$ and standard Einstein gravity
\beq
{\cal S}=\int d^4x\,\sqrt{-g} \left\{R-\frac{1}{2}\partial^\mu\Phi\partial_\mu\Phi-V(\Phi)\right\}
\eeq
The background metric is Friedman-Robertson-Walker
\beq
ds^2=-\bar{N}dt^2+a(t)^2\delta_{ij}dx^idx^j\,.
\eeq
Inhomogeneities are encoded in the perturbations of the scalar field and spatial metric, defined as
\beq
\Phi(t,\vc{x})={\phi}(t)+\varphi(t,\vc{x})
\eeq
\beq
\gamma_{ij}(t,\vc{x})=a(t)^2 e^{2\zeta(t,\vc{x})}\delta_{ij}dx^idx^j
\eeq
Furthermore, the lapse and shift components of the metric also receive perturbations: $N(t,\vc{x})=\bar{N}(t)+n(t,\vc{x})$ and $N_i=\partial_i\psi(t,\vc{x})$.
The lowest order manifestation of non-Gaussianity is a non-zero 3-point function for the perturbations determined by their cubic interactions. The standard way for computing it since the pioneering paper of Maldacena \cite{Maldacena:2002vr} has been through the use of a canonical Hamiltonian formalism. Such an approach requires a choice of gauge, the elimination of $n(t,\vc{x})$ and $N_i=\partial_i\psi(t,\vc{x})$ by solving the constraints and the computation of the second order Hamiltonian, determining the propagation of the perturbations, and the third order Hamiltonian determining the lowest order of interactions. One should note that since there are time derivative interactions, care is necessary in defining the interaction Hamiltonian. The computation has so far been performed in two specific gauges, the spatially flat gauge $\zeta=0$ and the uniform field gauge $\varphi=0$.

In this short paper the computation of non-Gaussianity from the cubic interactions is formulated using functional methods in a gauge invariant form which unifies the treatments existing in the literature. We construct a gauge invariant generating functional from which all n-point functions involving the cubic interactions can be calculated and show how to use it for computing a gauge invariant 3-point function from which the 3-point function of the curvature perturbation can be obtained. This illustrates a new way of organizing the calculations and brings out the special role played by the spatially flat gauge. In fact the interaction terms in this gauge coincide with the gauge invariant ones. We also point out how the explicit choice of a hypersurface $t={\rm const}$ in more conventional formulations breaks the apparent gauge invariance of the theory through the appearance of boundary terms which need to be included for the correct calculation of the curvature perturbation 3-point function.

\vspace{1cm}

It has been long known \cite{Mukhanov:1990me} that the quadratic action for scalar perturbations can be brought to the form
\beq
 {\cal S}_{2}  = \int d^4x\, \frac{1}{2}\bar{N}a^3\epsilon\Big[{\dot w}^2 -
\Big(\frac{\partial_iw}{a}\Big)^2\,\Big]
\label{2nd-w}
\eeq
where
\beq
w=\frac{1}{\sqrt{\epsilon}}\varphi - 2\zeta
\eeq
is the Sasaki-Mukhanov variable and
\beq
\epsilon \equiv  \frac{1}{4}\frac{\dot\phi^2}{H^2}\,.
\eeq
Note that we are using units in which $16\pi G=1$. To obtain (\ref{2nd-w}) one usually eliminates the perturbed lapse and shift by solving the linearized energy and momentum constraints for $n(t,\vc{x})$ and $\psi(t,\vc{x})$ \cite{Mukhanov:1990me}. It is possible to obtain (\ref{2nd-w}) without solving the constraints and retain the lapse and shift perturbations as non-propagating fields in (\ref{2nd-w}) \cite{Prokopec:2010be}, which is useful for treating interactions above order 3. In this note we focus on cubic interactions so this does not make a difference for our purposes here. The fact that the action for scalar perturbations can be reduced to (\ref{2nd-w}) indicates that with a single scalar field, there is a single scalar degree of freedom which propagates.

Let us now consider gauge transformations. A gauge transformation in GR can be thought of as a diffeomorphism on the perturbed spacetime which moves points along the integral curves of a vector field $\xi^\mu$: $x^\mu \rightarrow x^\mu+\xi^\mu$. As a result any geometrical quantity Q - be it a scalar, a vector or a tensor - transforms according to
\beq
Q\rightarrow e^{\mathcal{L}_\xi}Q
\label{gxfn}
\eeq
where $\mathcal{L}_\xi$ is the Lie derivative along the vector field $\xi$. Note that general gauge transformations are more complicated \cite{Bruni:1996im} but for our purpose one can always write (\ref{gxfn}) which suffices.

Under a gauge transformation $w$ is invariant to first order but suffers a change to second order (see eg \cite{Rigopoulos:2002mc})
\beq
w \rightarrow w + \left(\frac{\ddot{\phi}}{H\dot{\phi}}-\frac{\dot H}{H^2}\right)\left[H^2\xi^2+2\zeta H\xi\right]+\left[\left(\frac{\ddot{\phi}}{H\dot{\phi}}-\frac{\dot H}{H^2}\right)w+\frac{\dot{w}}{H}\right]H\xi \equiv w+\Delta_2^\xi w\,.
\eeq
For simplicity we have ignored terms that are unimportant on long wavelengths, as well as vectors and tensors. $\xi$ is the ``gauge parameter'' of time displacements: $t \rightarrow t+ \xi(t,\vc{x})$. Thus, under a gauge transformation, ${\cal S}_2$ contributes a cubic term
\beq\label{DS2}
{\cal S}_2[w] \rightarrow  {\cal S}_2[w] + \int {\rm d}^4\!x \,\frac{\delta {\cal S}_2[w]}{\delta w(x)} \Delta_2^\xi w(x)\,.
\eeq
However, since GR is diffeomeorphism invariant, the action for perturbations should be invariant under gauge transformations. This implies that the cubic action must contain an invariant piece plus terms that are proportional to the equations of motion which will counter the gauge variation (\ref{DS2}). Keeping in mind that $\zeta$ transforms like
\beq
\zeta \rightarrow \zeta+ H\xi\,,
\eeq
we see that the cubic action must necessarily take the form
\beq\label{cubic}
{\cal S}_{3} = \tilde{\cal S}_3[w] -\int {\rm d}^4\!x \,\frac{\delta {\cal S}_2[w]}{\delta w}\left[\left(\frac{\ddot{\phi}}{H\dot{\phi}}-\frac{\dot H}{H^2}\right)\left(\zeta^2+w\zeta\right)+\frac{\dot{w}}{H}\zeta\right]\,,
\eeq
where $\tilde{\cal S}_3[w]$ is invariant to this order and the term proportional to $\frac{\delta {\cal S}_2[w]}{\delta w}$ transforms appropriately so as to cancel the gauge variation of ${\cal S}_2$ . Furthermore, if we now define
\beq\label{W}
W=w-\left[\left(\frac{\ddot{\phi}}{H\dot{\phi}}-\frac{\dot H}{H^2}\right)\left(\zeta^2+w\zeta\right)+\frac{\dot{w}}{H}\zeta\right] \equiv w-F(w,\zeta)\,,
\eeq
which is gauge invariant to second order\footnote{This coincides with the gauge invariant variable given in \cite{Malik:2005cy} and the one that arises in the covariant formalism of \cite{Langlois:2010vx}}, the action is written as
\beq
{\cal S}[W]=\int d^4x\, \bar{N}a^3\left\{\frac{1}{2}\epsilon\left[{\dot W}^2 -
\Big(\frac{\partial_iW}{a}\Big)^2\right]\right\} +\tilde{\cal S}_3[W]\,.
\label{cubic-inv}
\eeq
This is invariant up to quartic terms which we here ignore. To obtain the explicit form for $\tilde{S}_{3}[W]$ we can work directly in the $\zeta=0$ gauge. The second term in (\ref{cubic}) is then identically zero and we have
\bea \label{tildeS}\tilde{\cal S}_{3}[W]\!\! &=& \!\!\int {\rm d}^3\!x{\rm d}t
\,\frac{\bar{N}a^3}{2} \Bigg\{
\left(\frac{3}{2}\epsilon^2-\frac{\epsilon^3}{4}\right)W\dot{W}^2
      +\epsilon^2W(\partial_i\dot{W})\left(\frac{\partial_i}{\nabla^2}\dot{W}\right)
+\frac{\epsilon^2}{2}W\left(\frac{\nabla W}{a}\right)^2
\label{cubic-h=0}\\
&&+\,\left(\epsilon^3-\frac{3}{2}\epsilon^2\eta\right)
     H W^2\dot{W}
+\left(-\frac{3}{2}\epsilon^3+\frac{\epsilon^2V_{,\phi\phi}}{2H^2}
+\frac{\epsilon^{3/2}V_{,\phi\phi\phi}}{3H^2}
 +\frac{\epsilon^2\eta^2}{2}\right)H^2 W^3
%\nonumber\\
%&&\hspace{2.3cm}
     +\frac{\epsilon^3}{4}\,W\left(\frac{\partial_i\partial_j}{\nabla^2}
  \dot{W}\right)\left(\frac{\partial_i\partial_j}{\nabla^2}\dot{W}\right)
 \Bigg\}
\,, \nonumber \eea
where the second line contains terms that are subleading in slow roll. In other words, after the replacement $\varphi/\sqrt{\epsilon}\rightarrow W$, the uniform curvature gauge \emph{directly provides the gauge invariant interactions} $\tilde{\cal S}_3$ which are the pieces of the action relevant for the gauge invariant computation of non-Gaussianity. All other cubic terms appearing in an arbitrary gauge can be eliminated by the ``field redefinition'' of the form (\ref{W}). Equivalently, working in an arbitrary gauge and eliminating $\zeta$ for $W$ should automatically remove all terms proportional to the linear equations of motion.

\vspace{1 cm}

Let us now consider the computation of non-gaussian n-point functions. Expectation values of any function of $W$ at background time $t$ can be obtained from the in-in generating functional
\beq\label{in-in}
Z[J_-,J_+]=\sum\limits_\alpha \langle \Omega,t_{-\infty}|\alpha,t_{\rm out}\rangle_{J_-} \langle \alpha,t_{\rm out}|\Omega,t_{-\infty}\rangle_{J_+}\,,
\eeq
where $ |\Omega,t_{-\infty}\rangle$ is the vacuum state defined at $t=-\infty$, $J_+$ and $J_-$ are two arbitrary independent sources. If $t$ is the time of interest for calculating an expectation value, then $t_{out}\geq t$. Let us take $t_{out} \rightarrow +\infty$. Such a choice is particularly convenient as will be seen shortly since it allows us to freely drop any boundary term, an operation that was crucial for obtaining the forms of the action in the above formulae. We will comment on the other possibility $t_{out}= t$ below.

Expressed in terms of a path integral the generating functional can be written as
\beq
Z[J_-,J_+]=\int[{\cal D}W_+{\cal D}W_-] \,
{\rm e}^{iS[W_+]\,-\,iS[W_-]\,+\,iJ_+W_+\,-\,iJ_-W_-}\delta(W_+(t_{\rm out}) - W_-(t_{\rm out}))
\,,
\eeq
The boundary conditions are specified in the usual way at $t \rightarrow -\infty$ ignoring the interactions and any non-linear field redefinitions. There is no boundary condition specified at $t \rightarrow +\infty$ apart from the equality $W_+(t_{\rm out})= W_-(t_{\rm out})$, enforced by the $\delta$-function in the path integral.
The in-in generating functional leads to a set of Feynman rules including four types of two-point functions. For the purposes of computing equal time tree-level expectation values two types suffice:
\beq
i{\rm \Delta}_{++}(x;x^\prime)
            =\langle\Omega|T W(x)W(x^\prime)|\Omega\rangle
\,,\qquad
i{\rm \Delta}_{+-}(x;x^\prime)
            =\langle\Omega|W(x^\prime)W(x)|\Omega\rangle
\,,
\label{w propagators}
\eeq
satisfying
\beq
  \Big(\!-\!\partial_t\epsilon a^3{\bar N}{\partial_t}
        +\bar N \epsilon a\nabla^2\Big)\,i{\rm \Delta}_{++}(x;x^\prime)
                          =\ i\delta^4(x-x^\prime)
\,,\qquad
   \Big(\!-\!\partial_t\epsilon a^3{\bar N}{\partial_t}
        +\bar N \epsilon a\nabla^2\Big)\,i{\rm \Delta}_{\pm}(x;x^\prime)
                          = 0
\,,
\label{w propagators:2}
\eeq
The gauge invariant three point function $\langle W(t,\vc{x})W(t,\vc{y})W(t,\vc{z})\rangle$ can then be calculated from two Feynman diagrams
\vskip 0.3cm
\beq\parbox{20mm}{
\begin{fmffile}{cubic1+}
\begin{fmfgraph*}(30,30)
\fmfleftn{i}{2}\fmfrightn{o}{1}
\fmflabel{+}{i1}
\fmflabel{+}{i2}
\fmflabel{+}{o1}
\fmf{plain}{i1,v1,i2}
\fmf{plain}{v1,o1}
\fmflabel{+}{v1}
\end{fmfgraph*}
\end{fmffile}}
+\qquad\parbox{20mm}{
\begin{fmffile}{cubic1-}
\begin{fmfgraph*}(30,30)
\fmfleftn{i}{2}\fmfrightn{o}{1}
\fmflabel{+}{i1}
\fmflabel{+}{i2}
\fmflabel{+}{o1}
\fmf{plain}{i1,v1,i2}
\fmf{plain}{v1,o1}
\fmflabel{-}{v1}
\end{fmfgraph*}
\end{fmffile}}
\eeq
\vskip 0.3cm
\noindent The vertex corresponds to the interactions in (\ref{cubic-inv}) and the coordinates of the internal point $(u,\vc{u})$ are integrated over from $-\infty$ to $+\infty$. The first diagram corresponds to three $i{\rm \Delta}_{++}$ propagators connecting spacetime points $(t,\vc{x})$, $(t,\vc{y})$ and $(t,\vc{z})$ to the internal point $(u,\vc{u})$. The second diagram corresponds to three $i{\rm \Delta}_{+-}$ propagators. Note that from the definitions of the propagators the integrals for $u>t$ cancel.

The gauge invariant correlator has different interpretations in different gauges. In the uniform curvature gauge $\zeta=0$ and $\langle W(t,\vc{x})W(t,\vc{y})W(t,\vc{z})\rangle = \frac{1}{\epsilon^{3/2}}\langle \varphi(t,\vc{x})\varphi(t,\vc{y})\varphi(t,\vc{z})\rangle$. On the comoving gauge, $\varphi=0$, it represents the correlation of the quantity
\beq
\bar{\zeta}=\zeta-\left[\frac{1}{2}\left(\frac{\ddot{\phi}}{H\dot{\phi}}-\frac{\dot H}{H^2}\right)\zeta^2+\frac{\dot{\zeta}}{H}\zeta\right]\,.
\label{redef}\eeq
This is just the field redefinition first discussed by Maldacena \cite{Maldacena:2002vr}. Inverting (\ref{redef}) we can compute $\langle \zeta(t,\vc{x})\zeta(t,\vc{y})\zeta(t,\vc{z})\rangle$ from the gauge invariant $\langle W(t,\vc{x})W(t,\vc{y})W(t,\vc{z})\rangle$ with the addition of disconnected parts $\langle W(t,\vc{x})W(t,\vc{y})\rangle\langle W(t,\vc{x})W(t,\vc{z})\rangle$ etc.

Would we obtain the same result using the original action (\ref{cubic}) without eliminating the terms proportional to $\delta{\cal S}_2[w]/{\delta w}$ by resorting to the gauge invariant $W$? We would in this formulation. Suppose we had used the naive Feynman rules read off from (\ref{cubic}). In the uniform curvature gauge the actions (\ref{cubic}) and (\ref{cubic-inv}) coincide so there is no issue there. In any other gauge there are extra vertices from the operator in $\frac{\delta {\cal S}_2[w]}{\delta w}$. When acting on an $i{\rm \Delta}_{+-}$ line they contribute nothing since the latter satisfies the linear equation of motion. However, when acting on an $i{\rm \Delta}_{++}$ line they contribute a delta function which eliminates the integrations over the internal point. The result is then equivalent to the ``field redefinition'' (\ref{redef}).

Before closing let us discuss the choice of $t_{\rm out}$ in (\ref{in-in}). All of the above considerations relied on sending $t_{\rm out}\rightarrow +\infty$ in (\ref{in-in}) which in turn allowed us to discard all temporal boundary terms. We were thus led to an invariant formulation where one can extract correlators of quantities attached to specific gauges, such as $\varphi$ or $\zeta$, from the gauge invariant action for $W$. How do these considerations tie in with the more widely used approach for such computations in the literature? In most works so far one equates $t_{\rm out}=t$ in the generating functional (\ref{in-in}) or the equivalent quantity in the Interaction Picture of perturbation theory for the quantum operators. In this formulation one is led to a set of diagrammatic rules where there are no external $i{\rm \Delta}_{++}$ lines: all external lines are on-shell. Thus, in such computations of the tree level 3-point function all diagrams of the form
\vskip 0.3cm
\beq\parbox{20mm}{
\begin{fmffile}{cubic1-}
\begin{fmfgraph*}(30,30)
\fmfleftn{i}{2}\fmfrightn{o}{1}
\fmflabel{+}{i1}
\fmflabel{+}{i2}
\fmflabel{+}{o1}
\fmf{plain}{i1,v1,i2}
\fmf{plain}{v1,o1}
\fmflabel{-}{v1}
\end{fmfgraph*}
\end{fmffile}}
\eeq
\vskip 0.3cm
\noindent where the integration over the internal coordinates go only up to time $t$. It would then appear that any terms proportional to $\delta{\cal S}_2[w]/{\delta w}$ contribute nothing to the 3-point function. A field redefinition removing such terms would seem unnecessary or, alarmingly, the 3-point function would be defined only up to an arbitrary disconnected part, corresponding to an arbitrary field redefinition.

Such reasoning ignores gauge (in)variance. We saw above that field redefinitions which remove terms proportional to $\delta{\cal S}_2[w]/{\delta w}$ correspond to gauge transformations. In the conventional formulation, the use of $t$ as an upper limit in the action integrals breaks the gauge invariance of the theory. A gauge transformation changes the fields in the theory, but the corresponding diffeomorphism \emph{also displaces any physical hypersurface labeled by background time $t$}. Thus, even for a diffeomorphism invariant theory such as GR, action integrals defined up to background time $t$ suffer a change due to the displacement of the boundary. In this case the action up to time t can be written as
\beq\label{t-action}
{\cal S}=\int\limits^{t} d^3x dt' \, \bar{N}a^3\frac{1}{2}\epsilon\left[{\dot W}^2 -
\Big(\frac{\partial_iW}{a}\Big)^2\right]+\tilde{\cal S}_3[W]
             +\int d^3x \,F[W(t),\zeta(t)]\dot{W}(t)\,,
\eeq
where $F[W,\zeta]$ was defined in (\ref{W}). The last term in (\ref{t-action}) can now be recognized as a gauge dependent term arising from the dependence of the boundary on the choice of gauge. In this formulation, where all external lines are on-shell, the bulk action provides an invariant contribution while the surface term contributes a gauge dependent piece. As can be deduced from (\ref{t-action}), the contribution of this boundary term to the 3-point function is equivalent to the field redefinitions mentioned above, giving equivalent results for quantities tied to a particular gauge (see also \cite{Arroja:2011yj,Burrage:2011hd} for discussions on this point). In this way of viewing things, the special role played by the uniform curvature gauge also becomes evident since the interactions in that gauge coincide with the invariant part of ${\cal S}_3$.

Let us finally note that one could extend the above reasoning to higher order terms in the action. In particular one can expect the existence of terms in the 4th order action of the form
\beq
{\cal S}_4 \supset \int {\rm d}^4\!x \, \frac{1}{2}\frac{\delta^2 {\cal S}_2[W]}{\delta W^2}\Delta^\xi_4 W%+\frac{\delta {\cal S}_3[W]}{\delta W}\Delta^\xi_3 W
\eeq
which could again be removed by defining a third order invariant variable etc. The procedure could be iterated to provide the invariant vertices of the theory, which would correspond to those obtained in the uniform curvature gauge. In such an approach the inclusion of non-propagating fields in ${\cal S}_2$ corresponding to the lapse $n$ and shift $\psi$ \cite{Prokopec:2010be} would be necessary.

{\it Acknowledgements}: The author wishes to thank Bjorn Garbrecht, Cristiano Germani, Tomislav Prokopec and Yuki Watanabe for useful discussions and suggestions. This work is supported by the Gottfried Wilhelm Leibniz programme
of the Deutsche Forschungsgemeinschaft.


\begin{thebibliography}{99}

%\cite{Guth:1980zm}
\bibitem{Guth:1980zm}
  A.~H.~Guth,
 ``The Inflationary Universe: A Possible Solution To The Horizon And Flatness
  Problems,''
  Phys.\ Rev.\  D {\bf 23} (1981) 347.
  %%CITATION = PHRVA,D23,347;%%

%\cite{Goldwirth:1991rj}
\bibitem{Goldwirth:1991rj}
  D.~S.~Goldwirth, T.~Piran,
  %``Initial conditions for inflation,''
  Phys.\ Rept.\  {\bf 214 } (1992)  223-291.

%\cite{Komatsu:2010fb}
\bibitem{Komatsu:2010fb}
  E.~Komatsu {\it et al.},
  ``Seven-Year Wilkinson Microwave Anisotropy Probe (WMAP) Observations:
  Cosmological Interpretation,''
  arXiv:1001.4538 [astro-ph.CO].
  %%CITATION = ARXIV:1001.4538;%%

%\cite{CQG}
 \bibitem{CQG}
 Class.\ Quant.\ Grav.\  {\bf 27 } (2010) 12: Focus section on non-linear and non-Gaussian cosmological perturbations.

%\cite{Maldacena:2002vr}
\bibitem{Maldacena:2002vr}
  J.~M.~Maldacena,
  ``Non-Gaussian features of primordial fluctuations in single field
  inflationary models,''
  JHEP {\bf 0305} (2003) 013
  [arXiv:astro-ph/0210603].
  %%CITATION = JHEPA,0305,013;%%

%\cite{Mukhanov:1990me}
\bibitem{Mukhanov:1990me}
  V.~F.~Mukhanov, H.~A.~Feldman, R.~H.~Brandenberger,
  %``Theory of cosmological perturbations. Part 1. Classical perturbations. Part 2. Quantum theory of perturbations. Part 3. Extensions,''
  Phys.\ Rept.\  {\bf 215 } (1992)  203-333.

%\cite{Prokopec:2010be}
\bibitem{Prokopec:2010be}
  T.~Prokopec, G.~Rigopoulos,
  %``Path Integral for Inflationary Perturbations,''
  Phys.\ Rev.\  {\bf D82 } (2010)  023529.
  [arXiv:1004.0882 [gr-qc]].

%\cite{Bruni:1996im}
\bibitem{Bruni:1996im}
  M.~Bruni, S.~Matarrese, S.~Mollerach, S.~Sonego,
  %``Perturbations of space-time: Gauge transformations and gauge invariance at second order and beyond,''
  Class.\ Quant.\ Grav.\  {\bf 14 } (1997)  2585-2606.
  [gr-qc/9609040].

%\cite{Rigopoulos:2002mc}
\bibitem{Rigopoulos:2002mc}
  G.~Rigopoulos,
  %``On second order gauge invariant perturbations in multi-field inflationary models,''
  Class.\ Quant.\ Grav.\  {\bf 21 } (2004)  1737-1754.
  [astro-ph/0212141].

%\cite{Malik:2005cy}
\bibitem{Malik:2005cy}
  K.~A.~Malik,
  %``Gauge-invariant perturbations at second order: Multiple scalar fields on
  %large scales,''
  JCAP {\bf 0511} (2005) 005
  [arXiv:astro-ph/0506532].
  %%CITATION = JCAPA,0511,005;%%

%\cite{Langlois:2010vx}
\bibitem{Langlois:2010vx}
  D.~Langlois, F.~Vernizzi,
  %``A geometrical approach to nonlinear perturbations in relativistic cosmology,''
  Class.\ Quant.\ Grav.\  {\bf 27 } (2010)  124007.
  [arXiv:1003.3270 [astro-ph.CO]].

%\cite{Arroja:2011yj}
\bibitem{Arroja:2011yj}
  F.~Arroja, T.~Tanaka,
  %``A note on the role of the boundary terms for the non-Gaussianity in k-inflation,''
[arXiv:1103.1102 [astro-ph.CO]].

%\cite{Burrage:2011hd}
\bibitem{Burrage:2011hd}
  C.~Burrage, R.~H.~Ribeiro, D.~Seery,
  %``Large slow-roll corrections to the bispectrum of noncanonical inflation,''
[arXiv:1103.4126 [astro-ph.CO]].


\end{thebibliography}
\end{document}